\documentclass[%
 aip,
 amsmath,amssymb,
 reprint,%
]{revtex4-1}

\usepackage{graphicx}
\usepackage{dcolumn}
\usepackage{bm}
\usepackage{gensymb}
\usepackage{amsmath}
\usepackage{float}

\usepackage[utf8]{inputenc}
\usepackage[T1]{fontenc}
\usepackage{mathptmx}
\usepackage{xcolor}

\begin{document}

\preprint{AIP/123-QED}

\title{A cryogenic ice setup to simulate carbon atom reactions in interstellar ices}

\author{D. Qasim}
\email{dqasim@strw.leidenuniv.nl.}
 \affiliation{Laboratory for Astrophysics, Leiden Observatory, Leiden University, PO Box 9513, NL--2300 RA Leiden, The Netherlands}
\author{M. J. A. Witlox}%
\affiliation{Fine Mechanical Department, Leiden Institute for Physics Research (LION), Niels Bohrweg 2, NL--2333 CA Leiden, The Netherlands}

\author{G. Fedoseev}%
\affiliation{Laboratory for Astrophysics, Leiden Observatory, Leiden University, PO Box 9513, NL--2300 RA Leiden, The Netherlands}

\author{K.-J. Chuang}%
\affiliation{Laboratory Astrophysics Group of the Max Planck
Institute for Astronomy at the Friedrich Schiller University Jena, Institute of Solid State Physics, Helmholtzweg 3, D--07743 Jena, Germany}

\author{T. Banu}%
\affiliation{Institute for Theoretical Chemistry, University of Stuttgart, D--70569 Stuttgart, Germany}

\author{S. A. Krasnokutski}%
\affiliation{Laboratory Astrophysics Group of the Max Planck
Institute for Astronomy at the Friedrich Schiller University Jena, Institute of Solid State Physics, Helmholtzweg 3, D--07743 Jena, Germany}

\author{S. Ioppolo}%
\affiliation{School of Electronic Engineering and Computer Science, Queen Mary University of London, Mile End Road,
London E1 4NS, UK}

\author{J. K{\"a}stner}%
\affiliation{Institute for Theoretical Chemistry, University of Stuttgart, D--70569 Stuttgart, Germany}

\author{E. F. van Dishoeck}%
\affiliation{Leiden Observatory, Leiden University, PO Box 9513, NL--2300 RA Leiden, The Netherlands}

\author{H. Linnartz}%
\affiliation{Laboratory for Astrophysics, Leiden Observatory, Leiden University, PO Box 9513, NL--2300 RA Leiden, The Netherlands}

\date{\today}

\begin{abstract}

The design, implementation, and performance of a customized carbon atom beam source for the purpose of investigating solid-state reaction routes in interstellar ices in molecular clouds are discussed. The source is integrated into an existing ultrahigh vacuum setup, SURFace REaction SImulation DEvice (SURFRESIDE$^{2}$), which extends this double atom (H/D, O, and N) beamline apparatus with a third atom (C) beamline to a unique system that is fully suited to explore complex organic molecule solid-state formation under representative interstellar cloud conditions. The parameter space for this system is discussed, which includes the flux of the carbon atoms hitting the ice sample, their temperature, and the potential impact of temperature on ice reactions. Much effort has been put into constraining the beam size to within the limits of the sample size with the aim to reduce carbon pollution inside the setup. How the C-atom beam performs is quantitatively studied through the example experiment, C + $^{18}$O$_2$, and supported by computationally-derived activation barriers. The potential for this source to study the solid-state formation of interstellar complex organic molecules through C-atom reactions is discussed. 

\end{abstract}

\maketitle

\section{\label{intro}INTRODUCTION}

Complex organic molecules (COMs; carbon and hydrogen containing molecules with at least 6 atoms) have been detected in the cold and lightless environments of prestellar and starless molecular cloud cores (i.e., in the dark interstellar regions which are shrouded by dust), in addition to other astrophysical environments.\citep{herbst2009complex, soma2018complex, bacmann2019cold} 3-carbon COMs have now been observed in star-forming regions towards both, high-mass \citep{belloche2013complex,belloche2016exploring} and low-mass\citep{lykke2017alma} sources. Astrochemical models generally assume that a majority of the detected COMs in such surroundings originate from radical-induced surface reactions, in which the radicals are of molecular form.\citep{garrod2006formation,vasyunin2017formation} This is supported by a series of recent laboratory and theoretical investigations of solid-state reactions, such as HCO, CH$_3$O and CH$_2$OH recombinations, in which the radicals are formed by addition and abstraction reactions within the CO hydrogenation route.\citep{chuang2015h, butscher2017radical,fedoseev2017formation, alvarez2018hydrogen, lamberts2019formation}

Another solid-state pathway that offers a route to larger COMs is through direct carbon atom chemistry. This route has been proposed in theoretical works, \citep{tielens1997circumstellar,giovannelli2001bridge,charnley2005pathways, charnley2009theoretical} applied to observational studies, \citep{requena2008galactic} and recently in astrochemical models. \citep{simoni2020sensitivity} Neutral atomic carbon derived from the gas-phase is one of the most abundant elements in space,\citep{phillips1981abundance,van1988photodissociation,herbst2005chemistry} and is primarily available during the early period of ice formation (i.e., before it reacts to form CO gas).\citep{van1998chemistry,hollenbach2008water,taquet2014multilayer} The laboratory study of C-atom chemistry under conditions representative for cold molecular clouds (i.e., ground state atomic carbon on 10 K surfaces in an ultrahigh vacuum (UHV) environment) has turned out to be very challenging, as it is experimentally difficult to produce an intense beam of largely ground state atomic carbon. This is a reason why there is little known regarding the role and relevance of C-atom addition reactions in solid-state astrochemical processes. Recent laboratory works have demonstrated how atomic carbon can react to form simple radicals\citep{krasnokutski2016ultra,henning2019experimental} and COMs\citep{krasnokutski2017low} within liquid helium droplets. The present work extends on this with the first ice system capable to study C-atom chemistry reactions in interstellar ice analogues. 

The focus here is on the design, implementation, and characterization of an atomic carbon source into an existing atomic beamline setup, SURFace REaction SImulation DEvice (SURFRESIDE$^{2}$),\cite{ioppolo2013surfreside2} which is dedicated to studying molecular cloud surface reactions. The experimental details of this setup are described elsewhere.\citep{ioppolo2013surfreside2} SURFRESIDE$^{2}$ has been used to show how H$_2$O, CO$_2$, and COMs can form under interstellar cloud conditions.\citep{ioppolo2010water,chuang2015h,fedoseev2017formation,qasim2019alcohols} The two available atomic beamlines currently permit the formation of a number of radicals, including H/D, N, O, OH, and \textrm{NH$_x$}. The addition of an atomic carbon source further extends the possibilities to study COM formation by the accretion of atoms and small radicals, which is representative of the low density phase of molecular clouds where atoms are not yet largely locked up into molecules.\citep{linnartz2015atom,boogert2015observations}

The original design of the atomic carbon source is found in the work by \citet{krasnokutski2014simple} and the source discussed in this article is a customized SUKO-A 40 from Dr. Eberl MBE-Komponenten GmbH (MBE), patent number DE 10 2014 009 755 A1. The design of the tantalum tube that is filled with graphite powder can be found in the works by \citet{krasnokutski2014simple} and \citet{albar2017atomic} Heating of the tube causes the carbon to sublimate and react with the tantalum to produce tantalum carbide, resulting in the conversion of molecular carbon into atomic carbon. Thus, the advantage of this source is that it essentially produces C-atoms rather than C$_x$ clusters (<1\% C$_2$ and C$_3$ molecules).\citep{krasnokutski2014simple} Additionally, carbon atoms are formed by thermal evaporation rather than `energetic' processing. Therefore, we expect the formation of only ground state C($^{3}$P) atoms with moderate kinetic energies. The implementation, design, and calibration measurements of the source are described in section~\ref{sect2}. Its performance, shown through example reactions which are considered relevant from an astronomical perspective and useful for calibration purposes, is presented in section~\ref{results}. The results are interpreted following computationally-calculated activation barriers that are briefly discussed. Section~\ref{sect3} discusses how this source can be used to investigate astrochemically relevant surface reactions, and how it can contribute to the science proposed with the upcoming James Webb Space Telescope (JWST). Section~\ref{sect4} lists the concluding remarks by summarizing the pros and cons of the new setup described here.

\section{SURFRESIDE$^{3}$ and atomic carbon source description}
\label{sect2}

\begin{figure*}[hbt!]
\center
\includegraphics[width=16cm]{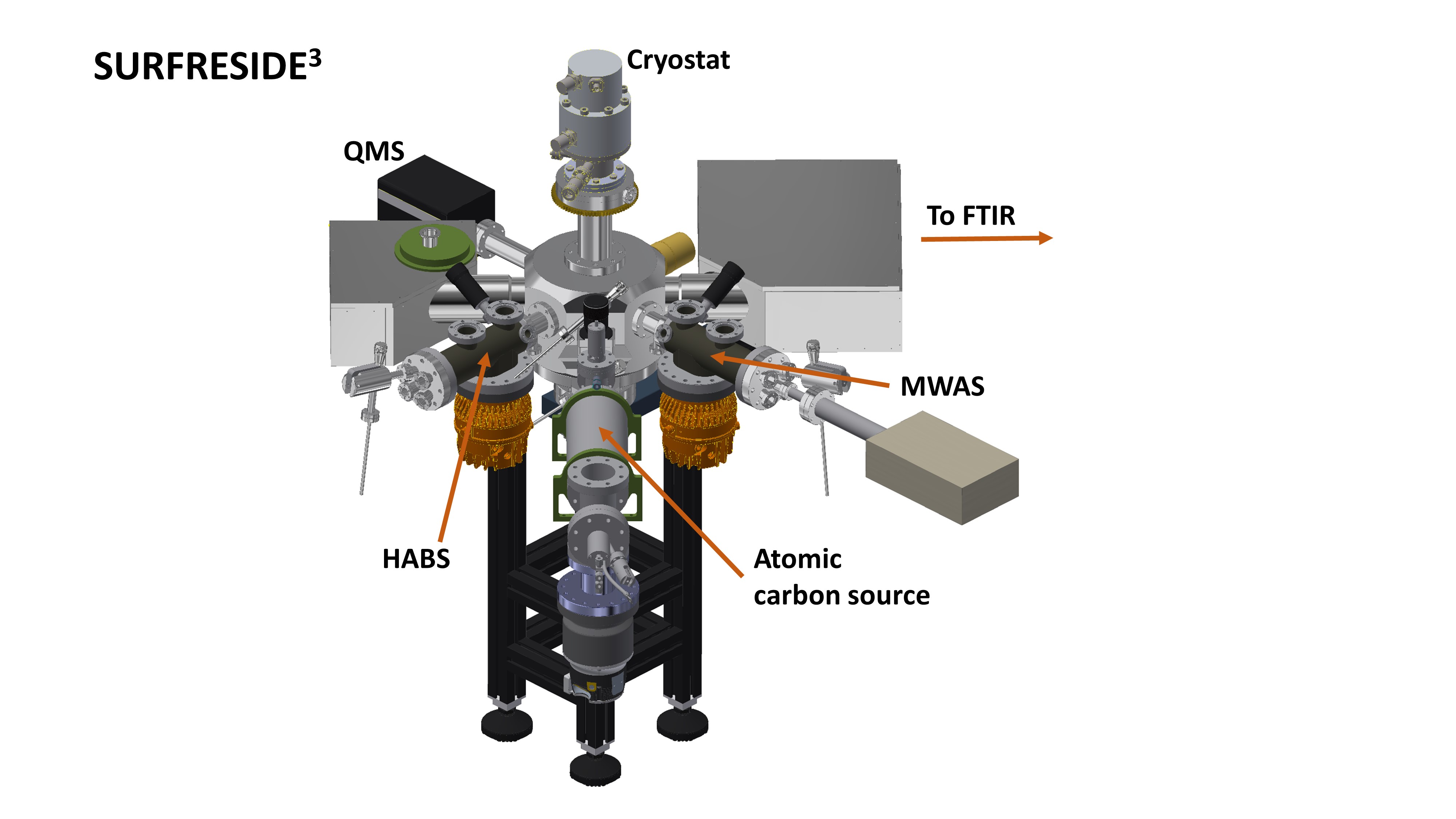}
\caption{\label{fig0.5} A three atomic beam line system, including the new C-atom source introduced here. The three atomic beam lines are capable of generating, H/D, N, O, and C-atoms, in addition to small radicals (e.g., OH, NH). It also contains two regular deposition lines. Both, pre-deposition and co-deposition experiments can be performed. RAIRS and TPD-QMS are used as diagnostic tools.}
\end{figure*}

The new C-atom beamline is implemented into an existing setup, SURFRESIDE$^{2}$, that has been described in detail before. \citep{ioppolo2013surfreside2} The extended system, SURFRESIDE$^{3}$, is shown in the 3D representation of Fig.~\ref{fig0.5}. This UHV system allows the growth of interstellar ice analogues on a sample surface for temperatures as low as 8 K using a closed-cycle helium cryostat. It comprises of three atomic beam lines. The HABS and MWAS beamlines have angles of 45$\degree$ to surface normal of the sample. The C-atom source is mounted in between the HABS and MWAS, and faces the plane of the surface of the ice substrate perpendicularly. The result of impacting H/D-atoms by the Hydrogen Atom Beam Source (HABS), \citep{tschersich1998formation,tschersich2000intensity,tschersich2008design} and/or H/D-, O-, N-atoms, and molecular radicals by the Microwave Atom Source (MWAS; Oxford Scientific Ltd.), and/or C-atoms by the new C-atom source is monitored using reflection absorption infrared spectroscopy (RAIRS) and/or temperature programmed desorption-quadrupole mass spectrometry (TPD-QMS). RAIRS allows monitoring of the formation of reaction products \textit{in situ}, as well as quantitative measurements of the newly formed products using a Fourier Transform Infrared Spectrometer (FTIR). TPD-QMS is complementary to RAIRS, as it exploits the desorption temperature, mass-to-charge (\emph{m/z}) value, and electron impact induced fragmentation pattern of the desorbed species to identify newly formed ice products. SURFRESIDE$^{3}$ is unique, as it allows to operate three different atomic beam lines simultaneously.

Ices are grown on the gold-plated substrate that is positioned vertically in the center of the main chamber of SURFRESIDE$^{3}$, which reaches a base pressure of $\sim$$3-4 \times 10^{-10}$ mbar at the start of each experiment. The surface is positioned such that it directly faces the C-atom source beam. A substrate temperature range of 8 -- 450 K is achieved by usage of a closed-cycle helium cryostat and resistive heating. The substrate temperature is probed by a silicon diode sensor that has an absolute accuracy of 0.5 K.

\subsection{Design of the C-atom line}

\begin{figure}[btp!]
\center
\includegraphics[width=9cm]{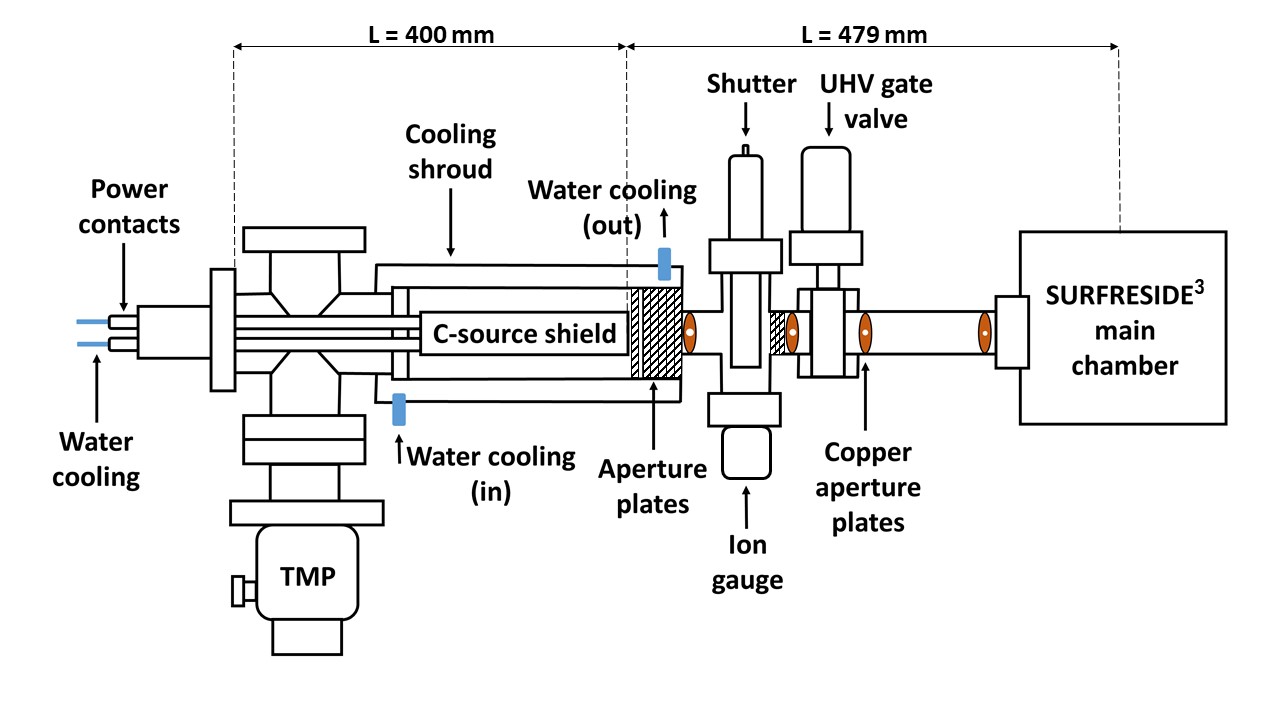}
\caption{\label{fig1} Side view schematic of the atomic carbon source vacuum chamber.}
\end{figure}

Fig.~\ref{fig1} illustrates the cross section of the C-atom source stainless steel chamber, which has a base pressure of $\sim$$2 \times 10^{-9}$ mbar when the source is at its standby current of 40 A. Two water-cooled power contacts are used to heat the source by means of a DC power supply that produces up to 1500 W (Delta Elektronika, SM 15-100).

The source is inserted into the vacuum chamber through a 4-way cross (CF 63). At the bottom of the cross hangs a turbomolecular pump attached to an adapter (Leybold 350i, 290 l/s for N$_2$, CF 100). As shown in Fig.~\ref{fig1}, the pump is placed behind the source to keep the pump at a distance from the carbon atom beam, as carbon deposits may stick and potentially harm the pump blades by applying weight to them. A water-cooled shroud is attached to the right flange of the 4-way CF 63 cross to prevent surrounding components from melting, as operating temperatures are around 2030\degree{C}. A tantalum shield is placed around the C-atom source for further protection. To the right of the shroud is a 4-way CF 40 cross piece. The top flange of the cross is attached to a rotary shutter. This shutter is situated in between the path of the emitting carbon atoms and the mini UHV gate valve in order to protect the gate valve from carbon build-up during ramping of the current. The gate valve is installed for the purpose of separating the C-atom source from the main chamber when necessary. At the bottom of the cross hangs a micro-ion gauge (Granville-Phillips, 355001-YG). Various sized aperture plates are installed to spatially restrict the carbon atom beam, where more details are found in section~\ref{beam}. 

Unlike the HABS and MWAS, the exit of the C-source does not have a nose-shaped quartz tube to help collisionally thermalize newly formed atoms before they impact the ices that are on top of the substrate; C-atoms have a much higher sticking coefficient and would coat the tube effectively with a carbon layer. This means that the impacting C-atoms carry the potential to induce thermal processing of the ice, which would not be representative of interstellar conditions. This is an important issue that has been addressed in more detail in the first science result with this new source; in \citet{qasim2020experimental} it was demonstrated that in a C + H + H$_2$O experiment -- combining the HABS and C-atom source -- the barrierless formation of CH$_4$ at 10 K predominantly follows a Langmuir-Hinshelwood mechanism (i.e., diffusing reactants thermalize prior to reaction on the surface). This suggests the likelihood of thermalization of the involved reactants, but it is not secured as to whether C-atoms thermalize with the substrate prior to reaction, as the formation of CH in the C + H + H$_2$O experiment may also proceed by Eley-Rideal (i.e., one reactant is not thermalized prior to reaction). For barrierless reactions, this is not relevant for qualitative studies, as such reactions will proceed regardless of the kinetic energy of the C-atoms. However, for reactions in which a barrier is involved, caution should be taken particularly for quantitative analysis, as the heat of the carbon atoms could open reaction pathways that are not accessible under typical interstellar conditions.

\subsection{Beam size calibration}
\label{beam}

\begin{figure}[hbt!]
\center
\includegraphics[width=6cm]{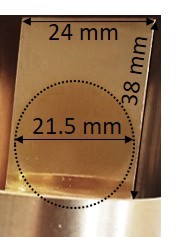}
\caption{\label{fig2} The darker circle, outlined by a dashed circle, is carbon deposit that resulted from the atomic beam on the gold surface of the substrate. The entire beam is on the flat surface, and the angle at which the picture is taken makes it appear that the lower part is clipped. A beam diameter of 21.5 mm is measured on a sample plate that has a size of 38 mm $\times$ 24 mm. Note that contrast has been added to the image to visualize the area with impacting C-atoms.}
\end{figure}

The beam size is measured on a gold-decorated substrate that has a width of 24 mm and length of 38 mm, and is positioned at a horizontal distance of 512 mm from the tantalum tube. Fig.~\ref{fig2} shows the resulting circular carbon atom beam diameter of 21.5 mm on the gold surface. The beam is narrowed by a combination of stacked oval-like aperture plates (produced by MBE) and circular copper aperture plates (produced in Leiden), as shown in Fig.~\ref{fig1}. From left to right, the aperture plates by MBE consist of one pyrolytic graphite (PG) plate with an orifice of 18$\times$6 mm, five tantalum plates with an orifice of 19$\times$7 mm, and three tantalum plates with an orifice of 20$\times$12 mm. The PG plate is placed directly after the source to allow the carbon to grow on it without introducing flakes, as thin graphite layers on thin metal parts sometimes create flakes. Multiple plates of the same aperture size are for the purpose of acting as radiation shields. The mean distance between the left-most and right-most aperture plate is 135 mm. The aperture sizes of the copper plates are 21 mm, 20 mm, 19 mm, and 18 mm, respectively, where the plate with the smallest aperture is placed closest to the substrate surface. From left to right, the distances between the copper plates are 126 mm, 35 mm, and 129 mm. The resulting beam size is optimized in that the majority of the atoms do not go past the sample plate, yet covers a large fraction of the substrate surface in order to have maximum overlap with the FTIR IR beam.     

\subsection{Temperature characterization of the graphite-filled tantalum tube}
\label{temp}

\begin{figure}[htb!]
\center
\includegraphics[width=9cm]{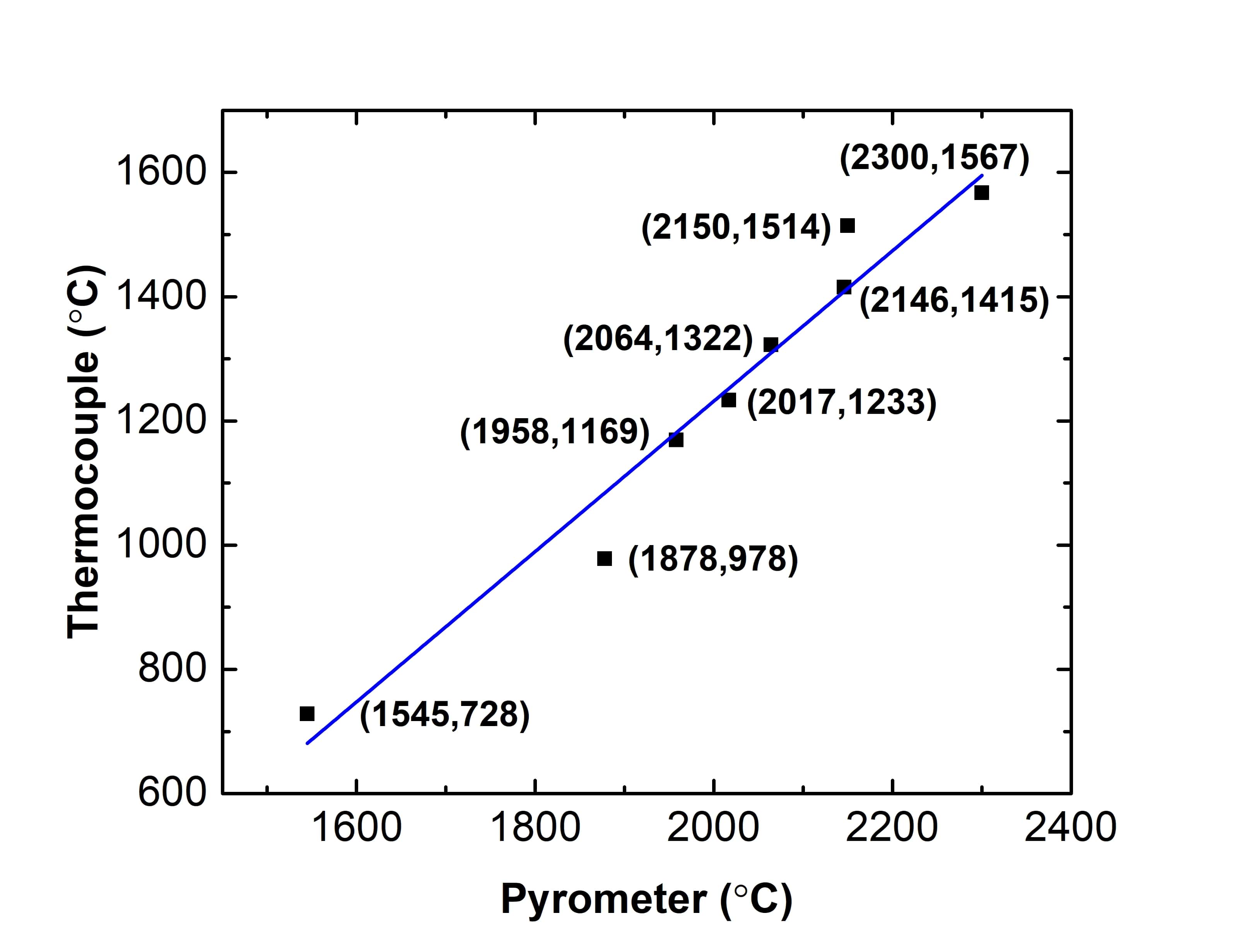}
\caption{\label{figgnew} A linear fit to the thermocouple versus pyrometer temperature values of the tantalum tube. The values from the pyrometer reflect the approximate C-atom temperatures, whereas the values from the thermocouple are lower due to the distance between the thermocouple and the heated tube.}
\end{figure}

An approximation for the temperature of the graphite-filled tantalum tube, and consequently of the emitting carbon atoms, is measured by a WRe alloy wire that is largely shielded with Al$_2$O$_3$ ceramic. A temperature controller (Eurotherm 2408) is primarily used to read out the temperature value. As the thermocouple is placed beneath the tube to protect it from melting, the measured value from the thermocouple is lower than the actual temperature of the heated tube. To know the actual temperature of the graphite-filled tantalum tube, and subsequently the emitting carbon atoms, a pyrometer is used (at MBE) in conjunction with the thermocouple, and the values are shown in Figure~\ref{figgnew}. Note that the C-atoms are assumed to be in thermal equilibrium with the tantalum tube, although in reality, their temperatures are lower, as the energy required to release the C-atoms from the tube (physisorption/chemisorption) is not taken into account. Interpolation of the values provides the approximate temperature of the carbon atoms for every thermocouple reading from 728 - 1567\degree{C}. These gas-phase temperature values are important for determining the flux of carbon atoms on the sample, as the flux is highly dependent on the filament temperature. Note that the pyrometer values are representative of ground state atomic carbon, as the amount of energy required to reach the C($^{1}$D) excited state is 14,665 K.\footnote{Kramida, A., Ralchenko, Yu., Reader, J., and NIST ASD Team (2018). NIST Atomic Spectra Database (ver. 5.6.1), [Online]. Available: https://physics.nist.gov/asd [2019, September 13]. National Institute of Standards and Technology, Gaithersburg, MD. DOI: https://doi.org/10.18434/T4W30F
}                

\subsection{C-atom flux calibration}

In order to characterize the reaction efficiencies in the ice, it is important to quantify the C-atom flux. In \citet{ioppolo2013surfreside2}, this was demonstrated for the HABS (H/D) and MWAS (H/D, N, O) fluxes. Here, the calibration procedure for the C-atom source is described through a quantitative analysis of various C + $^{18}$O$_2$ co-deposition experiments. RAIRS is used to probe and quantitate the reaction products. A wavelength range of 4000-750 cm$^{-1}$ with 1 cm$^{-1}$ spectral resolution is used. In section~\ref{results}, it will be explained that C + $^{18}$O$_2$ acts in a barrierless manner to ultimately form C$^{18}$O + $^{18}$O. $^{18}$O can then barrierlessly react with $^{18}$O$_2$ to form $^{18}$O$_3$.\citep{lin1982temperature,ioppolo2013surfreside2} Thus, in a very diluted C:$^{18}$O$_2$ matrix of $\sim$1:500, the $^{18}$O$_3$ abundance essentially reflects the C-atom abundance at the substrate. This aimed ratio of 1:500 is determined by estimating the C-atom flux using the C-atom flux values from MBE (application of the inverse-square law), and using the Langmuir approximation (1 L = $1.3 \times 10^{-6}$ mbar$\cdot$s = $1 \times 10^{15}$ molecules cm$^{-2}$) to estimate the $^{18}$O$_2$ flux. Note that an exact ratio of 1:500 is not critical in these specific experiments, as the main goal is to just create an overabundance of oxygen. Assuming a linear deposition rate, the flux of $^{18}$O$_3$, and thus atomic C, can be measured. To calculate the $^{18}$O$_3$ abundance, a modified Lambert-Beer equation is used, as employed in previous work.\citep{chuang2018reactive} A setup specific ozone band strength of $4.4 \times 10^{-17}$ cm molecule$^{-1}$ is applied, and derived by performance of an isothermal consumption experiment of O$_3$ by H-atoms at a deposition temperature of 50 K. The H-atom bombardment of O$_3$ ice at 50 K proceeds via an Eley-Rideal mechanism to form products, such as O$_2$, which desorbs upon formation. Thus, the rate of O$_3$ consumption by H-atoms is directly proportional to the H-atom flux (i.e., linear) until the surface is only partially covered by O$_3$. At that point the slope decreases, as the rate of O$_3$ consumption is hindered by the lack of O$_3$ at certain binding sites. Thus, the IR absorbance value of O$_3$ at the transition point is considered to represent a monolayer (ML; 1 ML = $1 \times 10^{15}$ molecules cm$^{-2}$) of O$_3$ ice, and is used to determine the band strength via the modified Lambert-Beer equation. A similar procedure was described in \citet{ioppolo2013surfreside2} to determine band strengths with SURFRESIDE$^{2}$.              

\begin{table}[htb!]
\caption{\label{table1.5} The $^{18}$O$_3$ (C-atom) flux values at various C-atom temperatures, measured from various C + $^{18}$O$_2$ co-deposition experiments. The C:$^{18}$O$_2$ ratio is aimed to be $\sim$1:500 in the experiments.}
\begin{ruledtabular}
\begin{tabular}{ccc}
Thermocouple & Deposition time & $^{18}$O$_3$ (C-atom) flux\\
(\degree{C}) & (s) & (cm$^{-2}$ s$^{-1}$) \\
\hline
1243 & 600 & $1 \times 10^{11}$ \\
1287 & 600 & $3 \times 10^{11}$ \\
1312$^{a}$ & 600 & $4 \times 10^{11}$ \\
1325$^{a}$ & 600 & $7 \times 10^{11}$ \\
1445 & 600 & $3 \times 10^{12}$ \\
1533 & 600 & $7 \times 10^{12}$ \\

\end{tabular}
\end{ruledtabular}
\footnotetext[1]{Measurement performed with another tantalum tube of the same design.}
\end{table}

The results are summarized in Table~\ref{table1.5}. An exponential curve can be fit to the values in the last column of Table~\ref{table1.5} in order to indirectly achieve C-atom flux values between thermocouple temperatures of 1243\degree{C} and 1533\degree{C}. The fluxes at the extremes are also measured by MBE for this particular source with a quartz crystal microbalance, and values of $1 \times 10^{11}$ cm$^{-2}$ s$^{-1}$ and $1 \times 10^{12}$ cm$^{-2}$ s$^{-1}$ are obtained for thermocouple temperatures of 1233\degree{C} and 1514\degree{C}, respectively. These do not deviate much from the values of $1 \times 10^{11}$ cm$^{-2}$ s$^{-1}$ (1233\degree{C}) and $5 \times 10^{12}$ cm$^{-2}$ s$^{-1}$ (1514\degree{C}) obtained with SURFRESIDE$^{3}$. The 5$\times$ deviation in flux at 1514\degree{C} can be due to a number of factors, such as the use of different vacuum chamber designs/geometries (including different pumping capacities), filament designs, and measurement tools. It is clear that for the use of this source in experiments for which flux values are needed, it is important to perform a setup specific calibration. Therefore, the method of measuring the $^{18}$O$_3$ abundance in an oxygen-rich C + $^{18}$O$_2$ co-deposition experiment to determine the C-flux should be repeated if changes are made. Note that the inverse-square law is applied in order to compare flux values to take properly into account the different distances involved in the two used experimental setups.

\section{Experimental and computational results}
\label{results}

The first results of carbon atom chemistry with the SUKO-A in SURFRESIDE$^{3}$ are presented below. The two experiments are listed in Table~\ref{table2}. These experiments are meant to test the performance of the source by conducting simple reactions that also are considered to be of astrochemical relevancy. 

\begin{table}[htb!]
\caption{\label{table2}Description of the performed experiments.}
\begin{ruledtabular}
\begin{tabular}{cccccccc}

No. & Exp. & T$_{sample}$ & T$_{C-atoms}$  & Flux$_{C-atoms}$ & Flux $^{18}$O$_3$ & Time \\
& & (K) & ($^{\circ}$C) & (cm$^{-2}$s$^{-1}$) & (cm$^{-2}$s$^{-1}$) & (s) \\
\hline

1 & C + $^{18}$O$_2$ & 10 & 1315 & $4 \times 10^{11}$ & $8 \times 10^{13}$ & 3000 & \\
2 & C + $^{18}$O$_2$ & 10 & 1315 & $4 \times 10^{11}$ & * & 3000 \\

\end{tabular}
\end{ruledtabular}
\footnotesize{Experiment 1 involves co-deposition, and experiment 2 involves pre-deposition. The C-atom fluxes are derived from interpolation of the flux values listed in Table~\ref{table1.5}. The Hertz-Knudsen equation\citep{Kolasinski2012} is used to determine molecular fluxes. T$_{\textrm{C-atoms}}$ refers to the temperature probed by the thermcouple. (*) refers to 10 Langmuirs (L).}
\end{table}

Isotopically enhanced gas, such as $^{18}$O$_2$ (Campro Scientific 97\%), is used to distinguish reaction products from possible contaminants. Other gases used are H$_2$ (Linde 5.0) and D$_2$ (Sigma-Aldrich 99.96\%). $^{18}$O$_2$ gas enters the main chamber of SURFRESIDE$^{3}$ through manually operated leak valves from turbomolecularly pumped dosing lines. Experiments proceed by either a pre-deposition or co-deposition manner. In the pre-deposition experiment, molecules are first deposited, followed by C-atom bombardment. In the co-deposition experiments, all species are deposited simultaneously. A major advantage of co-deposition is that product abundance is enhanced due to the constant replenishment of reactants in the ice upper layer. It is also more representative for interstellar processes.\citep{linnartz2015atom} Pre-deposition, on the other hand, allows monitoring of the kinetics of formation and consumption of products and reactants, as the initial abundance is known. This method is also preferred when layered ices have to be studied. More detailed information on the application of these two deposition methods can be found in \citet{schlemmer2014laboratory}. 

Relative molecular abundances are determined by using a modified Lambert-Beer equation, as done previously with the ozone abundance. The infrared band strength of C$^{18}$O (2086 cm$^{-1}$) used is $5.2 \times 10^{-17}$ cm molecule$^{-1}$.\citep{chuang2018reactive} For C$^{18}$O$_2$ (2308 cm$^{-1}$), a band strength of $4.2 \times 10^{-16}$ cm molecule$^{-1}$ is used. This value is obtained by multiplying the band strength of $7.6 \times 10^{-17}$ cm molecule$^{-1}$, which is from the work by \citet{bouilloud2015bibliographic} by a transmission-to-RAIR setup specific proportionality factor of 5.5, in which the band strength of CO from \citet{chuang2018reactive} is used. 





\begin{figure}[htb!]
\center
\includegraphics[width=7cm]{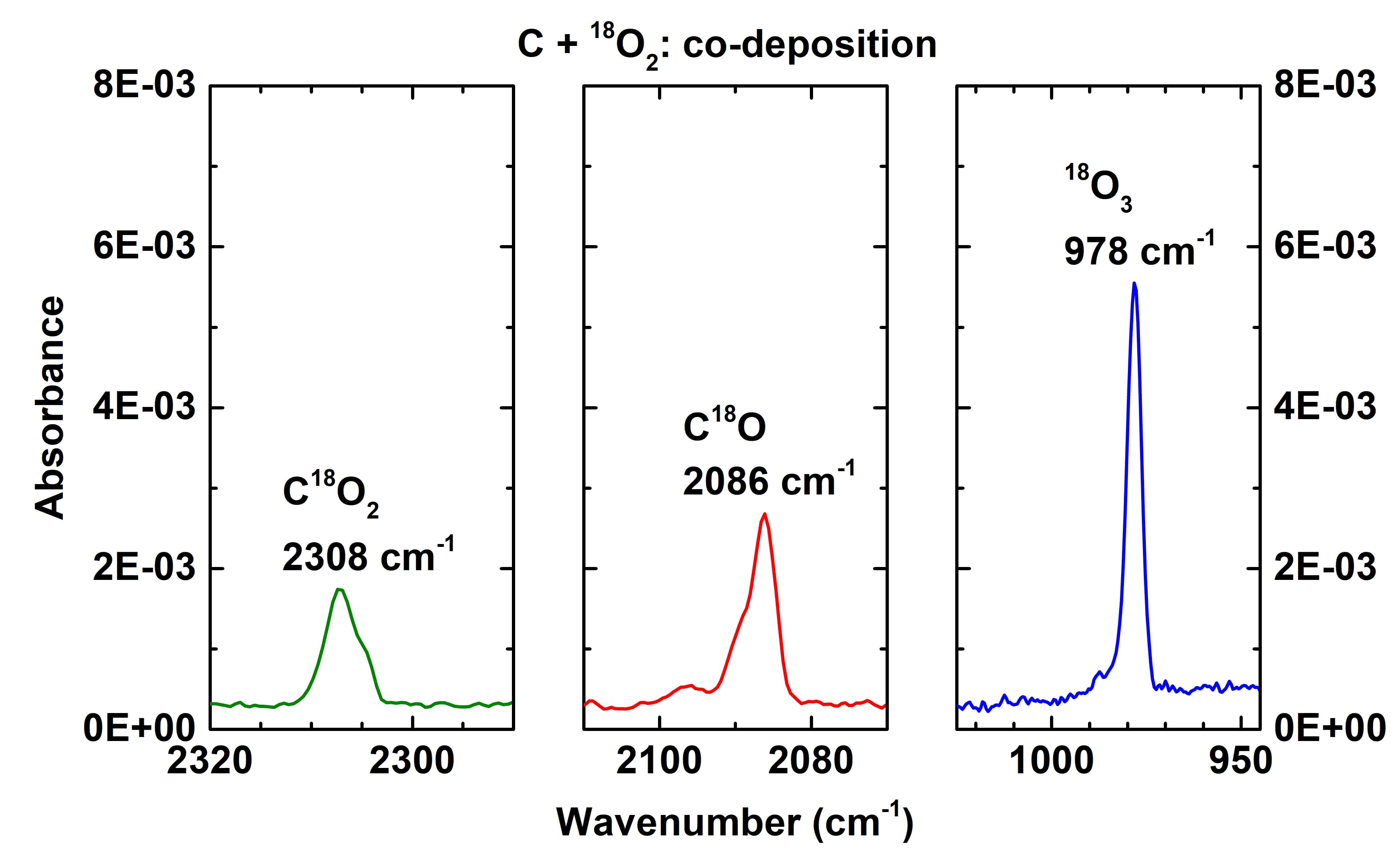}
\caption{\label{fig3} A RAIR spectrum acquired after co-deposition of atomic C and $^{18}$O$_2$ on a 10 K surface (exp. 1). The features of the reaction products, C$^{18}$O$_2$ (left), C$^{18}$O (middle), and $^{18}$O$_3$ (right), are highlighted. A C$^{18}$O:C$^{18}$O$_2$ abundance ratio of 12:1 is measured.}
\end{figure}

Fig.~\ref{fig3} features the IR signatures of the reaction products of the C + $^{18}$O$_2$ co-deposition experiment. Such products are C$^{18}$O$_2$, C$^{18}$O, and $^{18}$O$_3$. Particularly for the formation of C$^{18}$O$_2$ and C$^{18}$O, there may be more than one pathway to forming these species. Thus, for a more complete understanding of the C + O$_2$ reaction network, relevant computationally-derived activation and reaction energies are needed and are shown in Table~\ref{table3}. Connecting these energy values to the experimental results can delineate the product formation pathways. 

All the density functional theory (DFT) calculations are performed in a gas-phase model, which can be expected to cover the main effects, \citep{meisner2017atom} using Turbomole, \citep{TURBOMOLE} accessed via ChemShell.\citep{sherwood2003quasi,metz2014c} The unrestricted M06-2X functional\citep{zhao2008m06} is used in conjunction with the def2-TZVP\citep{weigend1998ri} basis set. The geometry optimizations are carried out at the same level of theory using the DL-Find\citep{kastner2009dl} optimizer interfaced via ChemShell. The connecting first-order saddle points (transition states) are obtained using the dimer method.\citep{henkelman1999dimer,kastner2008superlinearly} Numerical Hessians in DL-Find are used to characterize the optimized geometries as local minima or transition states. Reported energies include harmonic vibrational zero point energies. Intrinsic reaction coordinate (IRC)\citep{meisner2017comparison,hratchian2004accurate} calculations are performed to confirm the connections between the transition states and the local minima. Benchmark calculations are carried out using coupled-cluster level UCCSD(T)-F12,\citep{adler2007simple,knizia2009simplified} with a restricted Hartree-Fock reference and cc-PVTZ-F12 \citep{peterson2008systematically} basis set in Molpro.\citep{werner2012molpro}

\begin{figure}[htb!]
\center
\includegraphics[width=7cm]{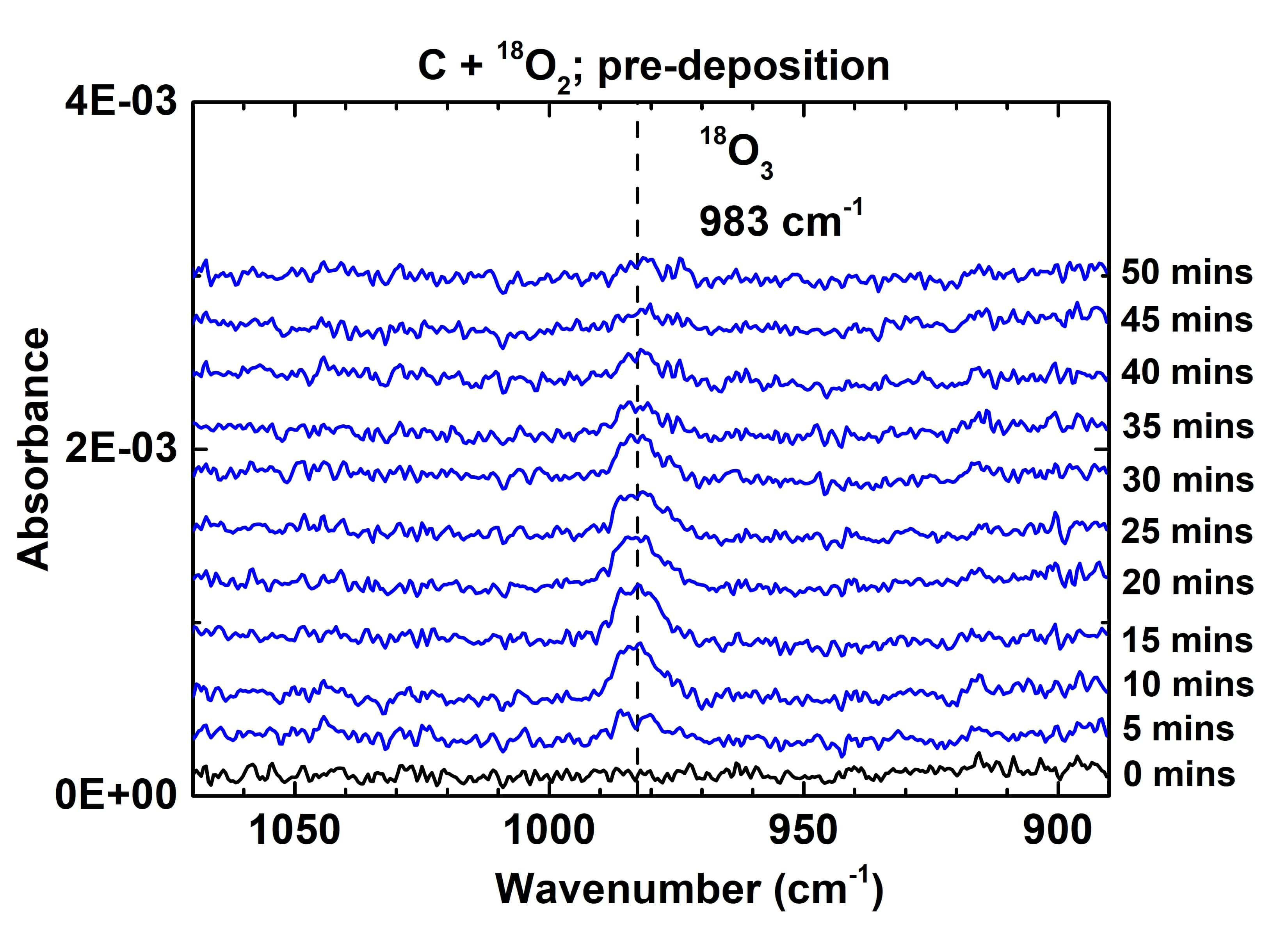}
\caption{\label{fig4} RAIR spectra acquired after pre-deposition of atomic C and $^{18}$O$_2$ on a 10 K surface (exp. 2). 10 L of $^{18}$O$_2$ is first deposited, followed by carbonation for 50 minutes. The increase and subsequent decrease of the $^{18}$O$_3$ band is highlighted. RAIR spectra are offset for clarity. }
\end{figure}

As demonstrated from the computational work, the reaction of C + O$_2$ barrierlessly leads to the intermediate, linear C-O-O. This process is exothermic by 410 kJ/mol in comparison to C + O$_2$ $\rightarrow$ CO + O, which has a reaction energy of -372 kJ/mol. In this context, it is noteworthy that both, C and O$_2$, are in their triplet ground states. As they combine to a singlet state, applying spin conservation results in the generated O-atom to also be in its excited singlet state. Thus, if energy is not dissipated into the ice, the overall process to CO formation is thought to be as fast as an actual barrierless reaction (decay of linear C-O-O to CO + O($^{1}$D)). CO can also be formed by the barrierless reaction of C + O$_3$ $\rightarrow$ CO + O$_2$, as listed in Table~\ref{table3}. In Fig.~\ref{fig4}, $^{18}$O$_3$ is formed by the barrierless reaction of $^{18}$O$_2$ + $^{18}$O, as shown by the increasing signal of $^{18}$O$_3$. Likely when $^{18}$O$_2$ becomes limited, $^{18}$O$_3$ starts to be consumed by C, as shown by the decreasing signal of $^{18}$O$_3$. Yet in the co-deposition experiment, it is unlikely that C$^{18}$O is formed from C + $^{18}$O$_3$, since the matrix of $^{18}$O$_2$ hinders the reaction between C and formed $^{18}$O$_3$.      

\begin{table}[htb!]
\caption{\label{table3} Activation and reaction energies for C + O$_2$, CO + O($^{3}$P), C + O$_3$, C + CO$_2$, and C$_2$O + CO calculated at the M06-2X/def2-TZVP level of theory. Additionally, a benchmark is performed with the CCSD(T)-F12/VTZ-F12 functional.} 
\begin{ruledtabular}
\begin{tabular}{cccc}
 Reaction & Product(s) & Activation energy & Reaction energy\\
 & & (kJ/mol) & (kJ/mol)\\
\hline

C + O$_2$\footnotemark[1] & CO + O($^{1}$D) & 0\footnotemark[3] & -372  \\
& CO$_2$\footnotemark[2] & - & -1106\\
CO + O($^{3}$P) & CO$_2$ & 25 & -527\\
C + O$_3$ & CO + O$_2$ & 0 & -981\\
C + CO$_2$ & CO + CO & 29 & -540\\
C$_2$O + CO & C$_3$O$_2$ & 30 & -346 

\end{tabular}
\end{ruledtabular}
\footnotetext[1]{Formation of the intermediate, linear C-O-O, is further discussed in the main text.}
\footnotetext[2]{Tentative (see main text for more details).}
\footnotetext[3]{Barrierless if energy from the formation of the linear C-O-O intermediate goes into the reaction (see main text for more details).}
\end{table}

If energy is dissipated into the ice, which is probable to occur as energy dissipation appears to occur within picoseconds,\citep{arasa2010molecular,fredon2017energy,qasim2019alcohols} linear C-O-O can decay to CO$_2$. However, it should be noted that a continuous reaction path to CO$_2$ is not found in the present computational simulations due to strong multireference character in the wave function, which is why CO$_2$ formation from C + O$_2$ is noted as tentative in Table~\ref{table3}. CO$_2$ can also be formed by CO + O($^{3}$P), albeit has a high barrier of 25 kJ/mol, if O($^{3}$P) is in fact formed. CO + O($^{1}$D) is barrierless to CO$_2$ formation.\citep{talbi2006interstellar} The abundance ratio of 12:1 for C$^{18}$O:C$^{18}$O$_2$ measured in experiment 1 shows that C$^{18}$O is the more favored product, and thus C$^{18}$O$_2$ formation is relatively inefficient under our experimental conditions.

The results of the C + $^{18}$O$_2$ experiments demonstrate that the C-atom source performs well under our astrochemically relevant experimental conditions to study solid-state reactions. Particularly, the available C-atom flux is sufficient to yield high quality spectra, which allows qualitative and quantitative analysis of the involved chemical pathways.

\section{Astrochemical implications}
\label{sect3}

 The new experiments described here are needed to understand how carbonaceous species, and particularly COMs, can be formed by carbon atom chemistry in the early phase of molecular cloud evolution. Such chemistry is expected to be most applicable to the H$_2$O-rich ice phase, which has a visual extinction ($A_{V}$) of 1.5 < $A_{V}$ < 3.\citep{boogert2015observations} At such cloud depths, atomic carbon is present, and becomes increasingly locked up in gas-phase CO at greater extinctions (> 3 $A_{V}$).\citep{van1998chemistry, boogert2015observations} This carbon then has the chance to react with atomic hydrogen on grain surfaces, for example, to form simple species such as CH$_4$.\citep{oberg2008c2d} The intermediate hydrocarbon radicals, \textrm{CH$_x$}, may also have the opportunity to react with other species in the H$_2$O-rich ice phase to form COMs. It should be noted that deeper into the cloud when CO freezes-out, C can also be formed by the dissociation of CO by cosmic-ray-induced processes.\citep{herbst2009complex,requena2008galactic}

The reaction of C + O$_2$ in interstellar molecular clouds may occur, as it is demonstrated to be a barrierless reaction, but not expected to be relatively frequent. Although interstellar O$_2$ ice has not been detected, O$_2$ has been detected on icy bodies, such as in the coma of comet 67P/Churyumov-Gerasimenko, and is thought to have primordial origin.\citep{bieler2015abundant,taquet2016primordial} As found in \citet{taquet2016primordial}, the distributions of the O$_2$ ice abundance at $A_{V}$ = 2, 4, 6, 8, and 10 are similar between the $A_{V}$ levels. Thus, the astrochemical timescales of relatively abundant C and O$_2$ should overlap. However, atoms such as H and O also barrierlessly react with O$_2$, and are at least an order of magnitude higher in abundance than C at the same timescales.\citep{hollenbach2008water,taquet2014multilayer,taquet2016primordial} Therefore, the reaction of C with O$_2$ in the translucent and dense phases of interstellar clouds is assumed to be minor for O$_2$ consumption. Nonetheless, the experimental and theoretical work presented here shows that atomic carbon and molecular oxygen can readily react if they would neighbor each other on a dust grain. The C + O$_2$ products, CO, and possibly O($^{3}$P) and CO$_2$, are unlikely to further react with each other in the laboratory or in the interstellar medium, as such reactions are associated with high activation barriers, as found in Table~\ref{table3}. This includes CO + O($^{3}$P) $\rightarrow$ CO$_2$ (25 kJ/mol), C + CO$_2$ $\rightarrow$ CO + CO (29 kJ/mol), and C$_2$O + CO $\rightarrow$ C$_3$O$_2$ (30 kJ/mol). However, the formation of O($^{1}$D) from C + O$_2$ may explain why CO$_2$ is formed starting from a `non-energetic' reaction, in which `non-energetic' refers to a radical-induced process that does not include an external energy source such as UV, cosmic rays, electrons, and/or simple heating of the ice. The reaction of C + O$_3$ is barrierless and thus may also occur in space. However, it is expected to be relatively infrequent for the same reasons as that for the C + O$_2$ reaction. 

The general relevance for the astrochemical community of using a C-atom source in a setup fully optimized to study atom addition/abstraction reactions in interstellar ice analogues is that it extends on reaction networks proposed before (e.g., \citet{charnley2005pathways}), but has not been investigated in the laboratory yet. It is expected that experimental investigations of solid-state C-atom chemistry will provide some of the missing fingerprints for how different carbon-bearing species are formed in interstellar ices. To date, the formation of solid-state COMs and other carbon-bearing molecules under interstellar relevant conditions is largely investigated through the combination of molecular radicals (e.g., HCO, C$_2$H$_3$, CH$_3$O), as this way to build the carbon backbone has been experimentally realized for some time. Although it is an important and relevant way to form solid-state carbon-containing species, it is likely not the explanation for the formation of all such species. This is in part due to the presence of atomic carbon in translucent and dense clouds. As atomic C is highly reactive, it may feasibly evolve into C$_x$H$_y$ structures. These structures can then react with other radicals to form alcohols and aldehydes, as shown in \citet{qasim2019alcohols} and \citet{qasim2019formation}. Alcohols, aldehydes, and other functional groups may also be formed starting from HCO + C \citep{charnley2005pathways,herbst2009complex} and/or CCO + H. Thus, this work will help understand the relative significance of radical recombination and direct C-atom addition reactions in various interstellar molecular cloud environments. For this, also astrochemical modeling will be needed, taking into account the available C-atom abundances in different astronomical environments. With the options SURFRESIDE$^{3}$ offers, it will become possible to provide information on possible reaction networks and reaction efficiencies and in full dependence of astronomically relevant temperatures. 

With the expected launch of the JWST in the near future, ice observations should become more prevalent, which also increases the necessity of laboratory C-atom reactions. To date, telescopes have suffered from telluric contamination (ground-based) and lacked the sensitivity to probe ice molecules more complex than methanol. The JWST will provide telluric free data, sensitivity and/or spectral resolution in the mid-IR that is orders of magnitude higher than encountered with previous telescopes such as the Infrared Space Observatory (ISO) or Spitzer,\citep{martinez2012mid} and perform ice mapping. These traits alone make the JWST a desired facility for ice observations, and preparations to search for COMs have already been carried out in the laboratory.\citep{van2018infrared} As possible COMs from C-atom chemistry are expected to be formed primarily in the H$_2$O-rich ice phase, and not in the CO-rich ice phase, C-atom chemistry experiments may provide insight into the origin of the COMs that will be targeted by the JWST. For example, a COM that is effectively formed by C-atom chemistry, but not by other studied pathways, suggests that the detected COM is formed in the polar phase of the cloud. Currently, not even methanol has been directly and exclusively detected in this phase in quiescent clouds (see Figure 7 of \citet{boogert2015observations}), and this is a consequence of the relatively low amount of ice at such low extinctions ($A_{V}$ < 3). Thus, this is the ideal time period to investigate C-atom chemistry, as the JWST may have the sensitivity to detect COMs directly in the polar phase that are formed starting from atomic C.  

\section{Conclusions}
\label{sect4}

For the first time, an atomic carbon source capable of producing fluxes in the low 10$^{11}$ -- high 10$^{12}$ cm$^{-2}$ s$^{-1}$ range is incorporated into a modified setup that is designed to study the `non-energetic' chemical processes of interstellar ice analogues. The source comes with new advantages: 1) An alternative way to investigate carbon chemistry in space along a principle that has not been studied so far. 2) A reliable and straightforward method to calibrate the C-atom flux in SURFRESIDE$^{3}$ is available. The flux is adequate to probe C-atom chemistry in SURFRESIDE$^{3}$, such as the reaction of C + $^{18}$O$_2$. The experimental results and computationally-derived activation barriers suggest that atomic carbon can react with O$_2$ and O$_3$ ice in interstellar molecular clouds, although more abundant species will effectively compete with C. 3) The beam size can be directly measured, which makes it achievable to operate the source without inducing hazardous carbon pollution into the vacuum system. The use of the source also comes with challenges to keep in mind: 1) The production of carbon layers on the sample surface is unavoidable in an experiment (i.e., all experiments take place on a carbonaceous surface). However, the layers observed have a negligible effect on the RAIR intensity. 2) The flux is highly dependent on the filament temperature, and the filament temperature steadily changes within an experiment partially due to the ongoing release of C-atoms. Thus, the longer the experiment, the greater the deviation of the flux between the start and the end of the experiment. 3) On average, the lifetime of a tube is around 14 hours at thermocouple temperatures of around 1300\degree{C}. This complicates experiments due to the necessary replacement of the tube, which can be expensive. 4) The extent of thermalization of the C-atoms to the temperature of the substrate is not fully secured yet, and therefore C-atom reactions involving activation barriers require caution, particularly if quantitative analysis is performed. Future studies will focus on developing a method to measure the extent at which C-atoms thermally equilibrate with the sample.    

With the positive performance of the modified setup, it is now possible to test what type of COMs can be formed by C-atom chemistry, primarily in a H$_2$O-rich ice, as these type of COMs are thought to be mixed primarily with H$_2$O (and also some CO). Such investigations overlap well with the expected launch of the JWST, which will have a sensitivity in the mid-IR that can possibly pick up signatures of COMs formed in low extinction ($A_{V}$ $\sim$ 2-3) environments directly -- something that has yet to be conducted with current observational facilities.

\begin{acknowledgments}
This research benefited from the joint financial support by the Dutch Astrochemistry Network II (DANII) and NOVA (the Netherlands Research School for Astronomy). Further support includes a VICI grant of NWO (the Netherlands Organization for Scientific Research) and an A-ERC grant 291141 CHEMPLAN. D.Q. acknowledges Andreas Jendrzey for many helpful discussions and Vianney Taquet for insightful feedback. The Leiden team specifically thanks Jiao He for on spot support, regular discussions and critical feedback. T.B. and J.K. acknowledge funding by the European Union's Horizon 2020 research and innovation programme (grant agreement No. 646717, TUNNELCHEM). S.I. recognizes the Royal Society for financial support and the Holland Research School for Molecular Chemistry (HRSMC) for a travel grant.
\end{acknowledgments}

\bibliography{aipsamp.bib}

\end{document}